\documentclass[prb,11pt]{revtex4-1}
\usepackage{geometry}                % See geometry.pdf to learn the layout options. There are lots.
\usepackage{graphicx}
\usepackage{amssymb}
\usepackage{amsthm}
\usepackage{amsmath}

\usepackage{mathtools}

\theoremstyle{definition}
\newtheorem{definition}{Definition}

\newcommand{\beq}{
\begin{eqnarray}
}
\newcommand{\eeq}{
\end{eqnarray}
}

\usepackage{epstopdf}
\begin{document}

\title{The Metric of Yang-Mills Orbit Space on the Lattice}
\author{M. Laufer}
\address{Department of Mathematics, CUNY Graduate Center, New York, NY, 10016-4309}

%\affiliation{No Affiliation}

\author{P. Orland}
\affiliation{Department of Natural Sciences, Baruch College, CUNY, New York, NY, 10010-5585 and Physics Program, CUNY Graduate Center, New York, NY, 10016-4309}
\thanks{P.O. supported in part by NSF grant no. PHY0855387 and a grant from PSC-CUNY}

\date{\today}                                           % Activate to display a given date or no date

%\subjclass[2010]{ Primary:   81T25 ; Secondary: 53Z05 }

\begin{abstract}
We find coordinates, the metric tensor, the inverse metric tensor and the 
Laplace-Beltrami operator for the orbit space of Hamiltonian  
${\rm SU}(2)$ gauge theory on a finite, rectangular lattice, with open boundary conditions. This is done using a complete axial gauge fixing. 
%The Gribov problem can be completely solved, with no remaining gauge ambiguities. 
%We briefly discuss non-generic points of this orbit space.
\end{abstract}

\maketitle

\section{Introduction}

There is no analytic method to determine the spectrum of the Hamiltonian
of the gauge theory of the strong interaction. There is a long-standing conjecture that there is a gap in this spectrum between the ground-state and first-excited-state energies. One strategy to illuminate this problem is to study the physical space of configurations 
\cite{singer}. These configurations are not choices of gauge field; rather they are gauge orbits. In this paper, we find coordinates on this space, eliminating gauge-fixing ambiguities \cite{gribov}. This is more difficult in three or more space-time dimensions than in two
\cite{1+1}. 

The space of gauge orbits is not a manifold, but an orbifold \cite{FSS}. The Hamiltonian of the gauge theory is a linear combination of the Laplace-Beltrami operator and a certain potential function 
on orbit space.

Determining some geometric quantities on orbit space (such as the Laplace-Beltrami operator, the Ricci curvature or the scalar curvature) require the 
evaluation of a trace. Such a trace does not exist without a regularization. Singer proposed 
zeta-function regularization for this purpose \cite{singer}. In this paper we will regularize with a lattice. In particular, we use Kogut and Susskind's  gauge theory, defined on a spatial lattice, but with continuous time \cite{k&s} (see Reference  \cite{creutz} for a derivation of the Kogut-Susskind formalism from Euclidean path-integral lattice gauge theory \cite{wilson}, with the 
transfer matrix). 

To make our treatment of gauge fixing simple, we break with the practice of using a toroidal lattice, using instead an open rectangular lattice.

The metric tensor on the space of gauge orbits can be understood as the projection operator
which vanishes on gauge transformations \cite{singer}, \cite{bab-via}. Another point of view is to regard orbit space as a metric space; the same metric 
tensor 
arises naturally in such a context \cite{orl-metric}. This projection operator 
is singular by definition; it acts on functionals of the gauge field, not the physical wave functionals of  gauge 
orbits. This is why it is desirable to find coordinates on orbit space. The metric between points of orbit space on the lattice was discussed in Reference \cite{kud-mor-orl}. Several papers have partly reduced the number of orbit-space coordinates in an axial gauge \cite{orl5}, but here all redundancies are completely eliminated. There are conically-singular points in the orbit-space orbifold, which arise from
gauge configurations invariant under a subgroup of the gauge group \cite{FSS}.

Our approach to finding coordinates and the metric on lattice orbit space makes it possible to find the Riemann, Ricci and scalar curvatures, at least in principle. A lower bound on the Ricci curvature implies a gap in the spectrum of the kinetic term of the Hamiltonian \cite{BochLichn}. The problem of calculating the curvature with the lattice regularization is more difficult than might be expected. This is because the metric tensor and inverse metric tensor need to be explicitly calculated. The lattice metric constructed in \cite{kud-mor-orl} was derived by taking an infimum over distances between elements of two orbits. This can then be coordinatized to give a metric tensor. Finding the inverse metric tensor is nontrivial, but possible. The inverse metric tensor is contained in the Laplace-Beltrami operator, so can be extracted once coordinates have been chosen. Determining the inverse metric tensor is the focus of this paper.

We should mention that there is another way of studying the space of configurations of 
$2+1$-dimensional gauge theories using holomorphic coordinates which appear very useful \cite{KJM}.  Some 
results similar to those in Reference \cite{KJM} were obtained in a simple formalism
\cite{giv}. 

This paper is organized as follows. Some 
definitions are given in Section 
\ref{sec:prelim}. Gauss' law and the definition of orbit space are given in Section \ref{sec:orbit}. In
particular, we discuss how gauge-equivalent gauge configurations are eliminated. Section 
\ref{subsec:last} is the heart of the paper, where the last step in the gauge fixing is 
done. To place the ideas in context, we review the metric on orbit space in 
Section \ref{sec:metric}. To make the coordinates on this space explicit, we introduce Euler angles for gauge fields in Section \ref{sec:coord}. In Section \ref{sec:imt}, we describe the form of the inverse metric tensor on a finite, $2$-dimensional rectangular lattice for gauge group ${\rm SU}(2)$. Finally, we summarize our results and discuss some avenues for further work in Section \ref{sec:concl}.

\section{Preliminaries}\label{sec:prelim}

\begin{definition}
The $D$-dimensional \emph{lattice} is the graph whose set of vertices is a subset of 
$\mathbb{Z}^D$, and whose edges connect each vertex to its nearest neighbors.\end{definition}

\begin{figure}
  \caption{The finite rectangular lattice in 2 dimensions.}
  \begin{center}
   \includegraphics[height=150mm]{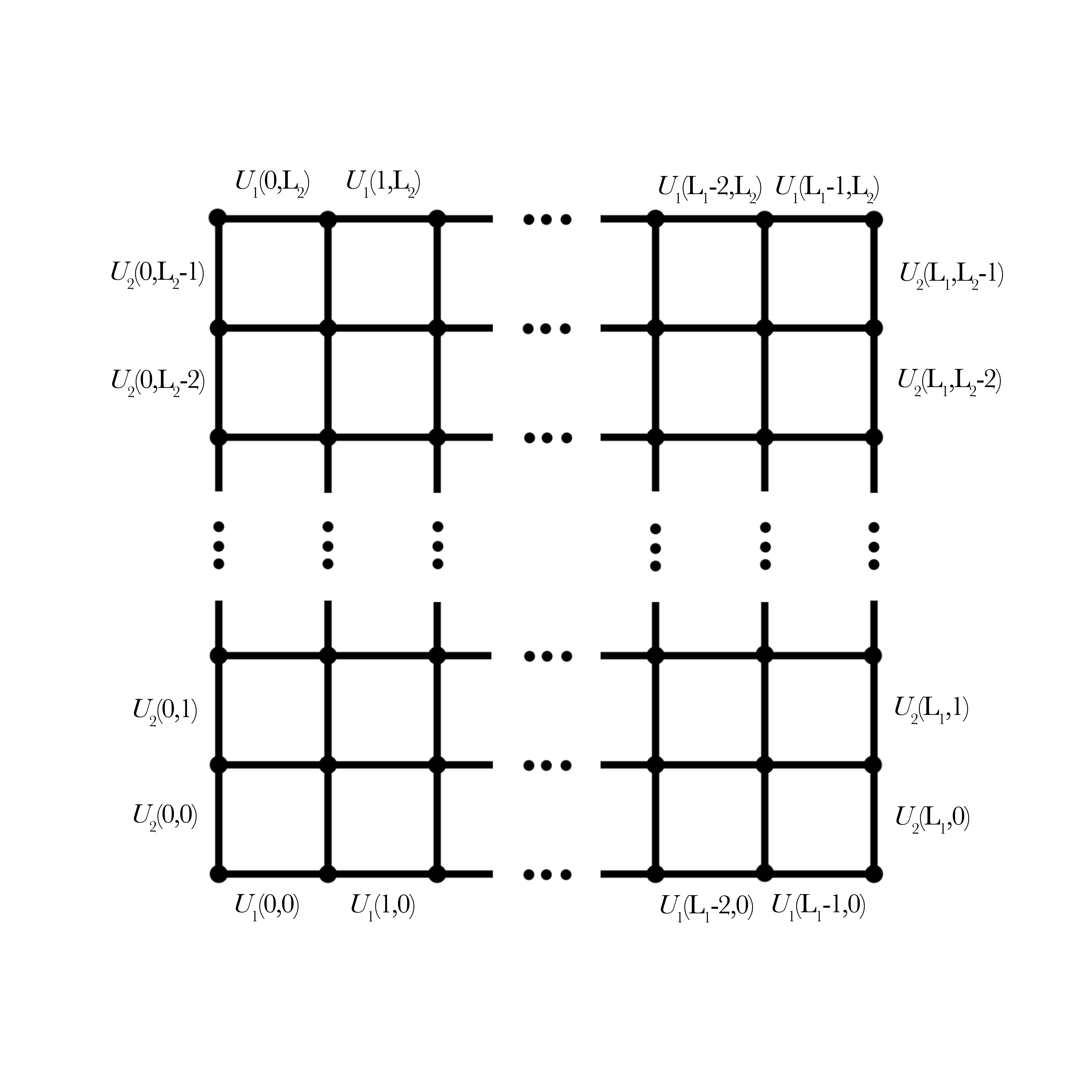}
     \end{center}
     \label{lattice}
\end{figure}

We will work with finite rectangular lattices. An example of such a lattice with $D=2$ is 
shown in Figure \ref{lattice}. 

The vertices of the lattice
are denoted by $\mathbf{x}\equiv(x_1,x_2,\dots,x_D)$. The numbers $x_{1}, \dots, x_{D}$ are integer multiples of the lattice spacing $a$, specifically $x_{j}=0,a,2a,\dots, L_{j}$, for $j=1,\dots,D$.
Let ${\hat 1}, \dots, {\hat D}$ be unit vectors in the positive $1-,\dots, D-$ directions, respectively. We denote the edge adjacent to the two vertices $\mathbf{x}$ and $\mathbf{x}+{\hat{\jmath}}$ 
by $(\mathbf{x},j)$, for each $j=1,\dots,D$. An element of ${\rm SU}(n)$ is assigned to each edge of the lattice. The ${\rm SU}(n)$ element
at the edge $(\mathbf{x},j)$ is denoted by $U_j(\mathbf{x})$, Henceforth we shall take $n=2$, for simplicity. In the lattice-gauge literature, the 
vertices are called \emph{sites} and the edges are called \emph{links}.

\begin{definition}
A \emph{wave function} is a complex-valued function of all of the variables $U_j(\mathbf{x})$ on all the edges. 
\end{definition}
\begin{definition}
\emph{Gauge state space} is the Hilbert space of square-integrable wave functions, with the inner product 
\beq
\langle\Psi\vert\Phi\rangle=\int\overline{\Psi(U_j(\mathbf{x}))}\Phi(U_j(\mathbf{x}))\prod_{\mathbf{x},j} 
dU_{j}(\mathbf{x}),
\nonumber
\eeq
where the integration measure on each edge is the Haar measure.
\end{definition}

We remark that only $n=2$ will be considered in any detail. We denote    
the basis vectors of $\mathfrak{su}(2)$ by $t_{1}$, $t_{2}$ and $t_{3}$, normalized 
by ${\rm Tr}\,t_{a}t_{b}=\delta_{ab}$.

A column vector $l_{j}(\mathbf{x})$ of the three differential operators, $[l_j(\mathbf{x})]_1$, $[l_j(\mathbf{x})]_2$
and $[l_j(\mathbf{x})]_3$,
is assigned to each edge $(\mathbf{x},j)$ of the lattice. We call these the {\em electric-field operators}.
They are defined by the 
commutation relations:
\begin{eqnarray}
[l_{j}(x)_{b} , l_{k}(y)_{c} ]=
{\rm i}{\sqrt 2}\delta_{x\,y}\delta_{j\,k} \;\epsilon^{bcd}
\;l_{j}(x)_{d} , \nonumber 
\end{eqnarray}
\begin{eqnarray}
[l_{j}(x)_{b}, U_{k}(y)]        =
-\delta_{x\,y}\delta_{j\,k}\; t_{b}\;U_{j}(x),
\nonumber
\end{eqnarray}
with all other commutators zero.

 \begin{definition}
 The \emph{Laplace-Beltrami operator} is
\beq
 -\Delta\equiv\sum_{\mathbb{Z}^D}\sum_{j=1}^D\sum_{j=1}^{n^2-1}{[l_j(\mathbf{x})]}_b^2.
 \nonumber
 \eeq
Its spectrum is unbounded and discrete. In $2$ dimensions this operator is:
\beq
-\Delta=\sum_{x_1=0}^{L_1-1}\sum_{x_2=0}^{L_2-1}\sum_{j=1}^2\sum_{b=1}^{3}{[l_j(x_1,x_2)]}_b^2.
\nonumber
\eeq
 \end{definition}

\begin{definition}
The \emph{covariant derivative} of $l_j(\mathbf{x})$ is 
\beq
\mathcal{D}_{j}l_j(\mathbf{x})\equiv\mathcal{D}_j(\mathbf{x})\cdot l_j(\mathbf{x})\equiv l_j(\mathbf{x})-(1-\delta_{0}^{x_j})U_j(\mathbf{x}-\hat{\jmath})l_j(\mathbf{x}-\hat{\jmath})U_j(\mathbf{x}-\hat{\jmath})^{-1}.
\label{CD1}
\eeq
\end{definition}

\noindent
The factor $(1-\delta_{0}^{x_j})$ is needed  because the lattice is finite and rectangular.

An element of the adjoint representation $\mathcal{R}_j(\mathbf{x})$, is assigned to 
$U_j(\mathbf{x})$ by
\beq
Ut_bU^{-1}\equiv\mathcal{R}t_b,
\nonumber
\eeq
where the arguments denoting the edge are implicit. Notice that $\mathcal{R}_j(\mathbf{x})$
lies in $SO(3)$. Hence \eqref{CD1} may be written
\beq
\mathcal{D}_{j}l_j(\mathbf{x})\equiv l_j(\mathbf{x})-(1-\delta_{0}^{x_j})\mathcal{R}_j(\mathbf{x}-\hat{\jmath})l_j(\mathbf{x}-\hat{\jmath}).
\label{CD2}
\eeq

\section{Orbit Space}\label{sec:orbit}

\begin{definition}
\emph{Gauss' law}  is 
\beq
\sum_j^D \mathcal{D}_{j}l_j(\mathbf{x})=0. \nonumber
\nonumber
\eeq
Gauss' law is imposed at every vertex.
\end{definition}

We denote by $\{U\}$
the collection of $U_{j}({\mathbf x})\in {\rm  SU}(2)$ for all the 
edges $({\mathbf x},j)$. The equivalence relation 
$\{U\} \simeq \{V\}$
between two lattice-gauge configurations $\{U\}$
and $\{V\}$
means that there is gauge transformation $\{K\}$, i.e. 
some collection $K({\mathbf x})\in {\rm SU}(2)$ at sites $x$ such that
\begin{eqnarray}
V_{j}({\mathbf x})\;=\;K({\mathbf x}+{\hat\jmath}a)^{-1}\;U_{j}({\mathbf x})\;K({\mathbf x})\;.
\nonumber
\end{eqnarray}
We will sometimes use the obvious notation 
$\{V\}=\{U\}^{\{K\}}$ for this expression. 

\begin{definition}
A \emph{gauge orbit} $u$ is 
an equivalence class of lattice-gauge configurations under the equivalence relation
$\simeq$, defined above. 
\end{definition}

\noindent Gauss' law is the statement that wave 
functions depend on orbits rather than
gauge configurations \cite{k&s} \cite{creutz}. To put coordinates on orbit space, we must first assign a unique element configuration for each equivalence class of 
gauge configurations. This is the procedure called {\em gauge fixing}.

A gauge transformation can easily be used to set the ${\rm SU}(2)$ elements on edges in the 
$1$-direction to unity. A further gauge transformation can then used to set the ${\rm SU}(2)$ elements 
on the edges in the $2$-direction for which $x_1=0$, and so on. We have thereby fixed the gauge on a maximal tree:
\beq
U_1(x_1, x_2, x_3, \ldots)=\mathbb{I}, \nonumber \\
U_2(0, x_2, x_3, \ldots)=\mathbb{I}, \nonumber \\
U_2(0, 0, x_3 \ldots)=\mathbb{I}, \nonumber \\
\vdots\;. \nonumber
\eeq
As this is done, we use Gauss' law to write the electric-field operators on the fixed edges
in terms of the 
electric-field operators on the unfixed edges. In $2$ dimensions this is
\beq
l_1(x_1,x_2)=-\sum_{y_1=0}^{x_1}\mathcal{D}_2 l_2(y_1,x_2), \label{2dgauss1}
\eeq
\beq
l_2(0,x_2)=-\sum_{y_2=0}^{x_2}\sum_{y_1=1}^{L_1}\mathcal{D}_2 l_2(y_1,y_2) .\label{2dgauss2}
\eeq
The procedure is similar for $D\geq 3$.

The Laplace-Beltrami operator for $D=2$ may now be rewritten as
\beq
\nonumber
-\Delta&=&\sum_{b=1}^3 \Bigg\lbrace \sum_{x_1=2}^{L_1} \sum_{x_2=0}^{L_2-1} [l_2(x_1,x_2)]^2 + \sum_{x_2=1}^{L_2-1} [l_2(1,x_2)]^2  \\ \nonumber
&& -\sum_{x_1=0}^{L_1-1} \sum_{x_2=0}^{L_2} \left[\sum_{y_1=0}^{x_1} \mathcal{D}_2 l_2(y_1,x_2)\right]^2 \label{LB} \\
&& -\sum_{x_2=0}^{L_2-1} \left[\sum_{y_2=0}^{x_2} \sum_{y_1=1}^{L_1} \mathcal{D}_2 l_2(y_1,y_2) \right]^2  \\
\nonumber
&& + \ [l_2(1,0)]^2 \Bigg\rbrace .\nonumber
\eeq

The gauge fixing in (\ref{2dgauss1}) and (\ref{2dgauss2})  is not yet complete. There are three remaining 
conditions to solve:
\beq
\sum_{y_2=0}^{L_2}\sum_{y_1=1}^{L_1} \mathcal{D}_2 {l}_2 (y_1,y_2)=0 .
\label{semifinalgauss}
\eeq

\subsection{Fixing the Last Edge}\label{subsec:last}

The remaining global condition (\ref{semifinalgauss}) can be solved by making a single
element of ${\rm SU}(2)$ (at one edge) diagonal. No further gauge fixing is then possible. For $D=2$, we chose to diagonalize $U_2(1,0)$. As a result $R_2(1,0)$ will also be diagonal. For this purpose, we rewrite (\ref{semifinalgauss}) as
\beq
-[\mathbb{I}-\mathcal{R}_2 (1,0)]l_2(1,0)=l_2(1,1)+\sum_{y_2=2}^{L_2}\mathcal{D}_2l_2(1,y_2)+\sum_{y_2=0}^{L_2}\sum_{y_1=2}^{L_1}\mathcal{D}_2l_2(y_1,y_2)\equiv\Xi.
\label{finalgauss}
\eeq

\section{The metric}\label{sec:metric}

The metric distance $\rho(u,v)$
between two gauge orbits $v$ and 
$v$ on the lattice 
is given by \cite{kud-mor-orl}
\begin{eqnarray}
\rho(u,v)^{2}
&=&N-\frac{1}{2}\inf_{\{K\} }\sum_{{\mathbf x}}\sum_{j=1}^{D} \left[
{\rm Tr}\; K({\mathbf x})V_{j}({\mathbf x})^{-1}K({\mathbf x}+{\hat\jmath}a)^{-1} U_{j}({\mathbf x}) \right. 
\nonumber \\
&+&\left. 
{\rm Tr} \;K({\mathbf x}+{\hat\jmath}a)V_{j}({\mathbf x})K({\mathbf x})^{-1}U_{j}({\mathbf x})^{-1} \right]
\;,\label{latmet}
\end{eqnarray}
where $\{U\}$ is any element of $u$ and $\{V\}$ 
is any element 
of $v$. This function of two orbits is gauge invariant. Furthermore, it is a metric \cite{kud-mor-orl}.

The partition function of a Wilson 
lattice gauge theory in $D+1$ dimensions, with discrete time $t$, is
\begin{eqnarray}
\prod_{{\mathbf x},t,\mu}\int dU_{\mu}({\mathbf x})\; e^{-S}\;, \label{wilson}
\end{eqnarray}
where the index $\mu$ runs from $0$ to 
$D$. The action $S$ may be split as 
\begin{eqnarray}
S=\frac{a^{D-2}}{a_{\rm t}g_{0}^{2}}
\sum_{t} {\mathcal L}_{\rm st}+\frac{a_{\rm t}a^{D-4}}{g_{0}^{2}}
\sum_{t}{\mathcal L}_{\rm ss}  \;, \nonumber
\end{eqnarray}
where $a_{\rm t}$ is the lattice spacing in the time direction, and
where 
${\mathcal L}_{\rm st}$ is the contribution of a space-time plaquette 
and ${\mathcal L}_{\rm ss}$ is the contribution
of a space-space plaquette. Explicitly 
\begin{eqnarray}
{\mathcal L}_{\rm st}\!&\!=\!&\!\frac{N}{2}
-\frac{1}{2}\sum_{{\mathbf x}}\sum_{j=1}^{D} \left[
{\rm Tr}\; U_{0}({\mathbf x},t)U_{j}({\mathbf x},t+a_{\rm t})^{-1}
U_{0}({\mathbf x}+{\hat\jmath}a, t)^{-1} U_{j}({\mathbf x},a) \right. \nonumber \\
\!&\!+\!&\!\left. {\rm Tr} \;U_{0}({\mathbf x}+{\hat\jmath}a,t)
U_{j}({\mathbf x},t+a_{\rm t})U_{0}({\mathbf x},t)^{-1}U_{j}({\mathbf x},t)^{-1} 
\right]\;,\label{space-time}
\end{eqnarray}
and
\begin{eqnarray}
{\mathcal L}_{\rm ss}\!&\!=\!&\!
\frac{N}{4}-\frac{1}{4}\sum_{{\mathbf x}}\sum_{j\neq k}\left[
{\rm Tr}\; U_{j}({\mathbf x},t)U_{k}({\mathbf x}+{\hat\jmath}a,t)
U_{j}({\mathbf x}+{\hat k}a,t)^{-1}U_{k}({\mathbf x},t)^{-1} 
\right. \nonumber \\
\!&\!+\!&\! \left. 
{\rm Tr}\; U_{k}({\mathbf x},t)U_{j}({\mathbf x}+{\hat k}a,t)
U_{k}({\mathbf x}+{\hat\jmath}a,t)^{-1}U_{j}({\mathbf x},t)^{-1} 
\right]\;.\label{space-space}
\end{eqnarray}
Note that the right-hand sides of (\ref{latmet}) and 
(\ref{space-time}) are very similar; if we
substitute for each $x$ and $j$
$U_{j}({\mathbf x},t)\rightarrow U_{j}({\mathbf x})$, 
$U_{j}({\mathbf x},t+a_{\rm t})\rightarrow V_{j}({\mathbf x})$, and
$U_{0}({\mathbf x},t)\rightarrow K({\mathbf x})$, into the right-hand side of 
(\ref{space-time}), and
take the infimum with respect to $K({\mathbf x})$, we obtain the 
lattice metric. Thus, by an appropriate
gauge fixing of the temporal gauge configuration $U_{0}({\mathbf x},t)$, we 
may replace ${\mathcal L}_{\rm st}$
by $\rho(u(t),u(t+a_{\rm t}))$, where $u(t)$ is the gauge orbit 
containing $\{U\}$ at time $t$ and 
$u(t+a_{\rm t})$ is the gauge orbit containing $\{U\}$ at 
time $t+a_{\rm t}$. Alternatively, if we
simply integrate out $U_{0}({\mathbf x},t)$, the dominant contribution 
to (\ref{wilson}) at weak coupling 
will come from this choice of $U_{0}({\mathbf x},t)$.

To see that (\ref{latmet}) is a metric, we note
that for any two orbits $u$ and $v$, 
$\rho(u,v)=\rho(v,u) \ge 0$, with
$\rho(u,v) = 0$ if and only if $u=v$. The only
remaining property we need is the triangle inequality, proved in Reference
\cite{kud-mor-orl}. As the proof is not hard, we repeat it below.

Notice that (\ref{latmet}) is the same as
\begin{eqnarray}
\rho(u,v)
&=&\inf_{\{K\} } \; I(\{U\},\{V\}^{\{K\}})
=\inf_{\{K\} } \; I(\{U\}^{\{K\}},\{V\}) \nonumber \\
&=&\inf_{\{K\},\{L\} } \; I(\{U\}^{\{K\}},\{V\}^{\{L\}})
\;, \label{latmet2}
\end{eqnarray}
where
\begin{eqnarray}
I(\{U\},\{V\})^{2}=
\frac{1}{2} \sum_{{\mathbf x}}\sum_{j=1}^{D} 
{\rm Tr}\; 
\left[ V_{j}({\mathbf x})- U_{j}({\mathbf x}) \right]^{\dagger}  
\left[ V_{j}({\mathbf x})- U_{j}({\mathbf x}) \right] \;. \label{latmet3}
\end{eqnarray}

Now for any three sets of matrices $\{U\}$ and $\{V\}$ $\{W\}$ we 
have that
\begin{eqnarray}
I(\{U \},\{V\})+I(\{V \},\{ W\}) \ge I(\{ U\},\{ W\})\;,
\nonumber 
\end{eqnarray}
which is a consequence of the triangle inequality of a vector 
space over the complex field (this
is formally true by (\ref{latmet3}), even if we are not dealing 
with special-unitary
matrices). Introducing gauge transformations $\{K\}$, $\{L\}$ 
and $\{M\}$, we have
\begin{eqnarray}
I(\{U \}^{\{K\}},\{V\}^{\{L\}})+I(\{V \}^{\{L\}},\{ W\}^{\{M\}}) 
\ge I(\{ U\}^{\{K\}},\{ W\}^{\{M\}})
\;,
\nonumber
\end{eqnarray}
which implies that
\begin{eqnarray}
I(\{U \}^{\{K\}},\{V\}^{\{L\}})+I(\{V \}^{\{L\}},\{ W\}^{\{M\}}) 
\ge \rho(u,w)
\;.
\nonumber
\end{eqnarray}
Taking the infimum of the left-hand side of this equation 
gives the triangle inequality 
\begin{eqnarray}
\rho(u,v)+\rho(v,w)\ge \rho(u,w)\;.
\label{triangle}
\end{eqnarray}

We next show that (\ref{latmet}) 
provides a Riemannian metric, except at conically-singular orbits \cite{FSS}. 

Let us substitute $U_{j}({\mathbf x})=e^{-i{\mathcal A}_{j}({\mathbf x})\cdot t}$, 
$V_{j}({\mathbf x})=e^{-i[
{\mathcal A}_{j}({\mathbf x})
+d{\mathcal A}_{j}({\mathbf x})
]\cdot t}$ and $K({\mathbf x})=e^{d\phi({\mathbf x})\cdot t}$ into (\ref{latmet}) 
and expand to second order in
$d{\mathcal A}_{j}({\mathbf x})$ and $d\phi({\mathbf x})$. The result is
\begin{eqnarray}
d\rho^{2}=\rho(u,v)^{2}=
\inf_{d\phi}
\sum_{{\mathbf x},j} \left\{ e_{j}({\mathbf x})_{\alpha}^{\;\;\;b}d{\mathcal A}_{j}({\mathbf x})^{\alpha}
+[-{\mathcal D}_{j}^{\dagger}d\phi({\mathbf x}+{\hat\jmath}a)]^{b} \right\}^{2} 
\;. \nonumber
\end{eqnarray}
The minimum of the sum on the right-hand side is unique. We 
find that $d\phi({\mathbf x})$ is  
\begin{eqnarray}
d\phi({\mathbf x})=\sum_{{\mathbf y},j}
({-{\mathcal D}^{\dagger}\cdot {\mathcal D}})^{-1}\delta_{\mathbf {xy}}\; 
{\mathcal D}_{j}\cdot e_{j}{\mathcal A}_{j}({\mathbf y})
\;, \nonumber
\end{eqnarray}
where
\begin{eqnarray}
e_{\alpha}^{\;\;\;a} t_{a}=-i U^{-1}\partial_{\alpha} U\;, \nonumber
\end{eqnarray}
(this is given explicitly by 
\begin{eqnarray}
e_{\alpha}^{\;\;\;a}  =
-i\left( \frac{{\mathbb I}- e^{i{\mathcal A}\cdot T}}{
{\mathcal A}\cdot T}\right)_{\alpha}^{\;\;\;a} 
\;,\nonumber
\end{eqnarray}
in canonical coordinates, where $T_{1,2,3}$ constitutes a basis of the adjoint representation of the Lie algebra) and where the Green's function 
$({-{\mathcal D}^{\dagger}\cdot {\mathcal D}})^{-1}\delta_{\mathbf {xy}}$ is 
uniquely determined by
the boundary conditions. 
This variational problem has the solution \cite{kud-mor-orl}
\begin{eqnarray}
d\rho^{2}=
G_{({\mathbf x},j,\alpha)({\mathbf y},k,\beta)}
d{\mathcal A}_{j}({\mathbf x})^{\alpha}
d{\mathcal A}_{k}({\mathbf y})^{\beta}
\;, \label{riemann1}
\end{eqnarray}
where we sum over lattice edges in our summation
convention and where the metric tensor is
\begin{eqnarray}
G_{({\mathbf x},j,\alpha)({\mathbf y},k,\beta)}=
e_{j}({\mathbf x})_{\alpha}^{\;\;\;b}
\left\{\delta_{\mathbf {xy}}\delta_{jk}\delta_{bc}-
\left[ (-{{\mathcal D}_{j}}^{\dagger})\frac{1}{{-{\mathcal D}^{\dagger}\cdot 
{\mathcal D}}}{\mathcal D}_{k}\right]_{bc}\delta_{{\mathbf {xy}}}
\right\}
e_{k}({\mathbf y})_{\beta}^{\;\;\;c}
\;. \label{riemann2}
\end{eqnarray}

Notice that the quantity in curled brackets in (\ref{riemann2}) is 
idempotent, hence it is a 
projection. In fact, the metric projects out
gauge transformations in inner products. To remove 
the zero eigenvalues, it is necessary to fix the gauge. The resulting induced metric is that on
all of orbit space, except at conical singularities. The set of these singularities is of measure zero, but it is an open question whether they have consequences for the Yang-Mills spectrum \cite{FSS}.

\section{Euler Angles}\label{sec:coord}

As in our previous discussion, we specialize to gauge group SU($2$). W introduce Euler coordinates at each edge:
\beq
U_{j}({\mathbf x})=e^{i\alpha_{j}({\mathbf x}){\sigma}_z}e^{i\beta_{j}({\mathbf x}){\sigma}_x}e^{i\theta_{j}({\mathbf x}){\sigma}_z},
\label{euler}
\eeq
where ${\sigma}_x, {\sigma}_y, {\sigma}_z$ are the Pauli matrices. The reason we use these coordinates (instead of the canonical coordinates of the previous section) is technical, not fundamental. It is easier to use Euler angles to perform the gauge fixing at the last edge.

In much of the discussion which follows, we denote the angles at the last edge $\alpha_{2}(1,0)$,
$\beta_{2}(1,0)$ and $\theta_{2}(1,0)$ by $\alpha$, $\beta$ and $\theta$, respectively and $U_{j}(1,0)$ by $U$.

A choice of basis vectors of $\mathfrak{su}(2)$  is
\beq
\mathcal{M}_{\gamma}^a \sigma_a \equiv -i \partial_{\gamma}U,  \label{Mdef}
\eeq
where $\gamma$ denotes $\alpha$, $\beta$ or $\theta$. From \eqref{euler}, we can find $\mathcal{M}_{\gamma}^a$:
\beq
\mathcal{M}=
\begin{pmatrix} 
\sin{2\alpha}\sin{2\beta}& -\cos{2\alpha}\sin{2\beta}&\cos{2\beta} \\ 
\\
\cos{2\alpha}&\sin{2\alpha}&0 \\
\\
0&0&1 
\end{pmatrix} . \label{M}
\eeq
This result can be used to express the electric-field operators in terms of derivatives of the coordinates:
\newline
\beq
\begin{pmatrix} l_1 \\ l_2 \\ l_3  \end{pmatrix}=\mathcal{M}^{-1}\begin{pmatrix}\partial_{\alpha}\\ \partial_{\beta} \\ \partial_{\theta} \end{pmatrix}. \nonumber
\eeq
Explicitly:
\beq
{[l_2(1,0)]}_1&=&\frac{\sin(2\alpha)}{\sin(2\beta)}\partial_{\alpha}-\cos(2\alpha)\partial_{\beta}-\frac{\cos(2\beta)\sin(2\alpha)}{\sin(2\beta)}\partial_{\theta} ,  \label{l's} \\
{[l_2(1,0)]}_2&=&\frac{\cos(2\alpha)}{\cos(2\beta)}\partial_{\alpha}+\sin(2\alpha)\partial_{\beta}-\sin(2\alpha)\partial_{\theta} , \nonumber \\
{[l_2(1,0)]}_3&=&\partial_{\theta}  \nonumber.
\eeq
From \eqref{euler}, and \eqref{l's}, $\mathcal{R}_j(x_1,x_2)$ can be explicitly calculated:
\beq
\;\;\mathcal{R}_j(x_1,x_2)=& \nonumber \\
\;\;&\begin{pmatrix} 
 \cos(2\beta)& -\cos(2\alpha)\sin(2\beta)&\sin(2\alpha) \sin(2\beta)\\ 
\\
 \sin(2\beta)\cos(2\theta)&
\begin{array}{c} -\sin(2\alpha)\sin(2\theta) \\ +\cos(2\alpha)\cos(2\beta)\cos(2\theta) \end{array} &
\begin{array}{c} -\cos(2\alpha)\sin(2\theta) \\ -\sin(2\alpha)\cos(2\beta)\cos(2\theta) \end{array}  \\
\\
\sin(2\beta)\sin(2\theta)& 
\begin{array}{c} \sin(2\alpha)\cos(2\theta) \\ + \cos(2\alpha)\cos(2\beta)\sin(2\theta) \end{array}&
\begin{array}{c} \cos(2\alpha)\cos(2\theta) \\  -\sin(2\alpha)\cos(2\beta)\sin(2\theta) \end{array}  
\end{pmatrix}  . \label{R}
\eeq
Diagonalizing $\mathcal{R}_2(1,0)$ by some $g\in {\rm SU}(3)$ yields the expression:
\beq
g^{-1}\mathcal{R}_2(1,0)g=
\begin{pmatrix}
1 & 0 & 0 \\
0 & k & 0 \\
0 & 0 & \frac{1}{k} \\
\end{pmatrix} , \label{diag1}
\eeq
where 
\beq
k&=& -\frac{1}{2}\left( (1-\cos(2\alpha-2\theta)[1+\cos(2\beta)-\cos(2\beta)]  \right.\nonumber \\
&+ &
\left.  \{4+[(1-\cos(2\alpha-2\theta)(1+\cos(2\beta)]-\cos(2\beta)\}^2)^{\frac{1}{2}}\right).
\label{diag2}
\eeq
Substitution of (\ref{diag1}) and (\ref{diag2}) into \eqref{finalgauss} gives two conditions, which 
allow the identification $\beta=\theta=\frac{\pi}{4}$. The gauge fixing is now complete.
We remove the ubiquitous factors of two in the remainder of this paper,  
by redefining
$\alpha\rightarrow 2\alpha$, $\beta\rightarrow 2\beta$, and $\theta\rightarrow 2\theta$.

Two of the derivatives on the $(1,0)$ edge may be written in terms of the third derivative, in addition to some of the angles on other edges:
\beq
\partial_{\theta}&=&\frac{\Xi_3}{1-\frac{1}{k}}, \label{theta} \\ 
\partial_{\beta}^{\star}&=& \frac{1}{\sin{\alpha}} \left(\sin{\alpha}\frac{\Xi_3}{1-\frac{1}{k}} +\frac{\Xi_2}{1-k} - \frac{\cos{\alpha}}{\cos{\beta}} \right) \label{beta},
\eeq
where $\Xi$ is defined in \eqref{finalgauss}. The asterisk on the derivative with respect to 
$\beta$ means that derivatives with respect to $\theta$ are replaced by the right-hand side of
\eqref{theta}.

The Laplace-Beltrami operator may now be written as:
\beq
-\Delta=-\Delta_{1}-\Delta_{2}-\Delta_{3}-\Delta_{4}-\Delta_{5}-\Delta_{6},\label{lb}
\eeq
where
\beq
-\Delta_{1}&=&
%\sum_{b=1}^3  
% \nonumber \\
%-\Delta_{2}&=&
%\sum_{b=1}^3  
\sum_{x_2=1}^{L_2} [l_2(1,x_2)]^2 +\sum_{x_1=2}^{L_1} \sum_{x_2=0}^{L_2} [l_2(x_1,x_2)]^2,   \nonumber \\
-\Delta_{2}&=& -
%\sum_{b=1}^3
\sum_{x_1=0}^{L_1-1} \sum_{x_2=0}^{L_2} \left[\sum_{y_1=0}^{x_1} \mathcal{D}_2 l_2(y_1,x_2)\right]^2, \nonumber  \\
-\Delta_{3}&=& -
%\sum_{b=1}^3  
\sum_{x_2=0}^{L_2-1} \left[\sum_{y_2=0}^{x_2} \sum_{y_1=1}^{L_1} \mathcal{D}_2 l_2(y_1,y_2) \right]^2,  \nonumber \\
-\Delta_{4}&=& \left(\sqrt{2} \sin{2\alpha} \ \partial_{\alpha} - \cos{2\alpha} \ \partial_{\beta}^{\star} - \sin{2\alpha} \ \frac{\Xi_3}{1-\frac{1}{k}}\right)^2,
\nonumber \\
-\Delta_{5}&=& \left(\sqrt{2} \cos{2\alpha} \ \partial_{\alpha} + \sin{2\alpha} \ \partial_{\beta}^{\star} - \sin{2\alpha}\frac{\Xi_3}{1-\frac{1}{k}}\right)^2, \nonumber \\
-\Delta_{6}&=&
\left(\frac{\Xi_3}{1-\frac{1}{k}}\right)^2, \nonumber  
\eeq
where the angle $\alpha\equiv \alpha_{2}(1,0)$ in $-\Delta_{4}$, $-\Delta_{5}$ and $-\Delta_{6}$ is the sole remaining 
coordinate specifying $U_{2}(1,0)$ (which is now diagonal).

A comparison of  with the standard form of the Laplace-Beltrami operator:
$-\Delta \equiv -\frac{1}{\sqrt{g}} \partial_{\mu} \sqrt{g} \ g^{\mu \nu} \partial_{\nu}, $
yields the inverse metric tensor.

\section{The Inverse Metric Tensor}\label{sec:imt}
The components of the inverse metric tensor may be read off by examining 
\eqref{lb}. Fortunately, the determinant of the metric is not needed to find these components. Finding 
any given component (that is, $g^{\mu \nu}$) is done by selecting the function between two partial derivatives of the associated coordinates and multiplying by the function in front. This is because each term of the Laplace-Beltrami operator in equation \eqref{lb} has the form 
$-\frac{1}{h_{\mu\nu}} \partial_{\mu} h_{\mu\nu} g^{\mu \nu} \partial_{\nu}$ (the square root of the determinant of the metric $\sqrt g$, automatically divides the
product $\prod_{\mu\nu}h_{\mu\nu}$).

To illustrate how the inverse metric tensor can be extracted, we give the example of the one-edge
Beltrami-Laplace operator $-\Delta_{\rm one-edge}= l^{2}$. From the expressions
\eqref{l's} for the components of $l$, we find
\beq
g^{\alpha\;\alpha }\!\!&\!\!=\!\!&\!\! \frac{1}{\sin^2\beta} ,\nonumber \\
g^{\alpha\;\beta}\!\! &\!\! =\!\! & \!\! 0 , \nonumber \\
g^{\alpha\;\theta}\!\! &\!\! =\!\! & \frac{\sin{\alpha}(\sin{(\beta - \alpha}))}{\sin^2 \beta} ,
\nonumber  \\
g^{\beta\;\beta}\!\!&\!\!=\!\!&\!\! 1 ,\nonumber \\
g^{\beta\;\theta}\!\!&\!\!=\!\!&\!\! \sin^2 \alpha + \frac{\sin{\alpha}\cos{\alpha}\cos{\beta}}{\sin{\beta}} ,\nonumber \\
g^{\theta\;\theta}\!\!&\!\!=\!\!&\!\!\frac{\sin^2 \alpha}{\sin^2 \beta} +1 , \nonumber
\eeq
%where the edge is implicit, i.e. the angles on each side of these equations are:
%$\alpha\equiv \alpha_{2}(x_{1},x_{2})$, $\beta\equiv \beta_{2}(x_{1},x_{2})$ and
%$\theta\equiv \theta_{2}(x_{1},x_{2})$. 
Nothing is new about this result, which is simply the inverse metric tensor of a three sphere.

Using \eqref{CD2}, \eqref{M}, and \eqref{R}, the components of $[\mathcal{D}_2 l_2(y_1,x_2)]_b$ 
(which are in $-\Delta_{2}$ and $-\Delta_{3}$) reduce to:
\beq
[\mathcal{D}_2 l_2(y_1,x_2)]_{1}&=&
\frac{\sin\alpha}{\sin\beta}\partial_{\alpha
%_{2}(y_{1}, x_{2} )
}-\cos\alpha\partial_{\beta
%_{2}(y_{1},x_{2})
}-\frac{\sin\alpha\cos\beta}{\sin\beta}\partial_{\theta
%_{2}(y_1,x_2)
}  \label{b1} \\ 
&+&\left(\cos^2\alpha-\frac{\sin\alpha\cos\beta}{\sin\beta}\right)\partial_{\alpha_{2}(y_1,x_2-1)} \nonumber \\
&+&\cos\alpha(\cos\beta+\sin\alpha\sin\beta)\partial_{\beta_{2}(y_1,x_2-1)} \nonumber \\	
&+&\sin\alpha\left(\frac{\cos^2\beta}{\sin\beta}+\cos\alpha\sin\beta-\sin\beta\right)
\partial_{\theta_{2}(y_1,x_2-1)} ,\nonumber 
\eeq
\beq
&&\!\!\!\!\!\!\!\!\!\!\!\!\!\!\![\mathcal{D}_2 l_2(y_1,x_2)]_{2}  \label{b2} \\
&=&\frac{\cos\alpha}{\sin\beta}\partial_{\alpha
%_{2}(y_1,x_2)
}+
\sin\alpha\partial_{\beta
%_{2}(y_1,x_2)
}+\sin\alpha\partial_{\theta
%_{2}(y_1,x_2)
}   \nonumber \\
&+&\frac{\sin\alpha\cos\alpha\sin\theta-\cos^2\alpha\cos\beta\cos\theta-\sin\alpha\cos\theta\sin\beta}{\sin\beta}\partial_{\alpha_{2}(y_1,x_2-1)} \nonumber \\
&+&(\sin^2\alpha\sin\theta-\sin\alpha\cos\alpha\cos\beta\cos\theta+\cos\alpha\sin\beta\cos\theta)\partial_{\beta_{2}(y_1,x_2-1)} \nonumber \\
&+&(\sin\alpha\cos\beta\cos\theta+\sin^2\alpha\sin\theta+\cos\alpha\sin\theta \nonumber \\
 &&  \ \ \ \ \ \ \ \ \ \ \ \ \ \ \ \ \ \   -\sin\alpha\cos\beta\cos\theta-\sin\alpha\cos\alpha\cos\beta\cos\theta)\partial_{\theta_{2}(y_1,x_2-1)}, \nonumber 
\eeq
\beq
&&\!\!\!\!\!\!\!\!\!\!\!\![\mathcal{D}_2 l_2(y_1,x_2)]_{3} \label{b3}\\
&=&\partial_{\theta
%_{2}(y_1,x_2)
} 
-\left (\sin\alpha\sin\theta+\frac{\sin\alpha\cos\alpha\cos\theta}{\sin\beta}+\frac{\cos^2\alpha\cos\beta\sin\theta}{\sin\beta}\right)\partial_{\alpha_{2}(y_1,x_2-1)} \nonumber \\
&+&(\cos\alpha\sin\beta\sin\theta-\sin^2\alpha\cos\theta-\sin\alpha\cos\alpha\cos\beta\sin\theta)\partial_{\beta_{2}(y_1,x_2-1)} \nonumber \\
&+&(\sin\alpha\cos\beta\sin\theta-\sin^2\alpha\cos\theta-\sin\alpha\cos\alpha\cos\beta\sin\theta \nonumber \\
&& \ \ \ \ \ \ \ \ \ \ \ \ \ \ \ \ \ \ 
+\cos\alpha\cos\theta-\sin\alpha\cos\beta\sin\theta)\partial_{\theta_{2}(y_1,x_2-1)} ,\nonumber
\eeq
where the edge direction and adjacent vertex are only indicated explicitly for coordinates other than 
$\alpha\equiv\alpha_{2}{(y_1,x_2)}$, $\beta\equiv\beta_{2}{(y_1,x_2)}$, 
$\theta\equiv\theta_{2}{(y_1,x_2)}$.

In $-\Delta_{2}$ we sum over spatial dimensions after squaring, but only in the $1$-direction. These terms are merely a local term coupled with an adjoint term from the edge below. These can be constructed similarly by multiplying the associated pieces, and then summing.

The term $-\Delta_{3}$ is somewhat more complicated, as the sums run in both directions. When constructing the contribution from this term for a given component of the metric tensor, it must be noted that there will be overlap from $\mathcal{D}_2l_2(y_1,y_2+1)$ with $\mathcal{D}_2l_2(y_1,y_2)$, except on the boundary. The form of these terms is $(l(x_{\gamma})-\mathcal{R} l(x_{\gamma})) (l(y_{\xi})-\mathcal{R} l(y_{\xi}))$, and can also be constructed similarly to the above.

The last three terms $-\Delta_{4}$, $-\Delta_{5}$ and $-\Delta_{6}$ contain many pieces. They have many more combinations than than the above, because they each contain sums over most of the lattice. Only the third component of the vector is taken, however, so summing of components is not required in their construction. 

This completely defines the inverse metric tensor for the finite rectangular lattice with $D=2$, working over ${\rm SU}(2)$. Similar methods work for $D\geq 3$. We believe it is possible to generalize to
the gauge group ${\rm SU}(n)$. Angular coordinates are considerably more complicated for $n>2$, however.

\section{Conclusions}\label{sec:concl}

To summarize, we have explicitly found the metric tensor for ${\rm SU}(2)$ on the lattice with open boundary conditions, and determined the inverse metric tensor.  It is noteworthy that the gauge-fixing problem only becomes
complicated when fixing the last edge.

The methodology used here can be generalized to construct results for higher-dimensional lattices and  and other gauge groups. There are no marked differences for $D\ge 3$. Generalizing the results to gauge group ${\rm SU}(n)$ is cumbersome, but the strategy is the same as for 
${\rm SU}(2)$; this is under investigation.

It should be possible to study orbit-space geodesics in our coordinates. Any geodesic in the full space of lattice gauge fields is described by the real parameter $t$ through
\beq
U_{j}({\mathbf x}; t)=\exp{\rm i}\,\tau({\mathbf x},j) \,t , \label{geod}
\eeq
where $\tau({\mathbf x},j)$ is an arbitrary chosen element of the Lie algebra chosen for each edge $({\mathbf x},j)$. The
geodesics in the completely-fixed axial gauge are obtained by gauge-fixing (\ref{geod}) according to the prescription given in this paper.

Finally, we believe that a detailed study of the set of conically-singular points \cite{FSS} in the lattice formulation of gauge theory should be very fruitful. This set is of measure zero, but that 
does not mean it has no significance. Presumably the Riemann curvature diverges at 
such points (even on the lattice). An interesting question is whether this divergence
can be regularized in a sensible and physically-meaningful way.

\section{Acknowledgements}
M.L. would like to thank David Stone and Tony Phillips for discussions. P.O. would like to thank both the Niels Bohr International Academy and the Kavli Institute for Theoretical Physics for their hospitality during which some of this work was done.

\bibliographystyle{plain}
\bibliography{ref}

\end{document}